\documentstyle[aps]{revtex}
\begin{document} 
\vskip 2truecm
\title{Cutting-Decimation Renormalization for
diffusive and vibrational dynamics on fractals}

\author{Raffaella Burioni\footnote{E-mail:
burioni@almite.mi.infn.it}}

\address{Dipartimento di Fisica and INFN,
Universit\`a di Milano, via Celoria 16, 20133 Milano, Italy and
INFM, Unit\`a di Parma}
\author{Davide Cassi\footnote{E-mail: cassi@vaxpr.pr.infn.it} and Sofia
Regina\footnote{E-mail: regina@vaxpr.pr.infn.it}}
\address{Dipartimento di Fisica,  INFM and INFN, Viale 
delle Scienze, 43100 Parma, Italy }
\maketitle
\vskip 1truecm
UPRF-97-11
\vskip 1truecm

\begin{abstract} 
Recently, we pointed out that on a class on non exactly decimable
fractals two different parameters are required to describe
diffusive and vibrational dynamics.
This phenomenon we call {\it dynamical dimension splitting} is
related to the lack of exact decimation invariance for these structures,
which turn out to be invariant under a more complex cutting-decimation
transform.
In this paper we study in details the dynamical dimension 
splitting on these fractals analyzing the mathematical properties
of the cutting-decimation transform. Our results clarify
how the splitting arises from the cutting transform and show
that the dynamical dimension degeneration is a very peculiar consequence of
exact decimability. 

\end{abstract}
\newpage
\section{Introduction}
Fractal dynamics is a field of primary importance in the study of
physical phenomena in real non-crystalline systems. 
The analytical study of dynamical properties of fractals requires 
the introduction of simplified model structures, which are supposed
to have the same universal properties of the real ones. However,
as a matter of fact, analytical results are usually 
restricted to exactly decimable fractals, where one can apply
the powerful techniques of exact renormalization group.
Exactly decimable fractals are indeed very peculiar structures, 
characterized by strong 
restrictions on their topology, which are far from being general and 
representative of all fractals.
Nevertheless, due to the availability of analytical results, their behavior 
was usually considered as typical.
This is the case for the spectral dimension of a fractal.
Such a parameter, the spectral dimension $\widetilde d$, was defined by the
asymptotic law \cite{aeo}:
\begin{equation}
\rho(\omega)\sim \omega^{{\widetilde d}-1}
\label{1a}
\end{equation}
where $\rho (\omega) $ is the density of harmonic vibrational modes with
frequency $\omega$, or by:
\begin{equation}
P_{ii}(t)\sim  t^{-{{\widetilde d}\over 2}}
\label{2a}
\end{equation}
where $P_{ii}(t)$ is the probability of returning to the starting site $i$ after
$t$ steps ( $t \rightarrow \infty$ ) for a random walker and
the exponent $\widetilde d/2$ is independent of the starting point \cite{hhw}.
In the original definition, (\ref{1a}) and (\ref{2a}) were supposed to be
equivalent by scaling arguments and in all calculations made on
exactly decimable fractals they have been found to be always equal.
However it has been shown \cite{prl} that from a mathematical
point of view it is not possible to conclude that the two asymptotic behaviors 
of (\ref{1a}) and (\ref{2a}) are equivalent.
Therefore, one should distinguish between the {\it diffusive spectral dimension}
defined by (\ref{2a}) and the {\it vibrational spectral dimension} defined by
(\ref{1a}), the former being relevant for local quantities and the latter for
bulk or average quantities.
In the following we will show that the coincidence of the two spectral 
dimensions is typical of exactly decimable fractals while more 
general structures present dynamical dimension splitting \cite{r+d+s}.
Indeed for exactly decimable fractals $\rho(\omega)$ and $P_{ii}(t)$ 
have the same transformation properties under renormalization and this leads
to the so called {\it spectral dimension degeneration}. 
Here we will consider the case of a class of non exactly decimable fractals
(called $NT_D$) showing that (\ref{1a}) and (\ref{2a}) can be calculated 
applying a generalized renormalization transformation, different
from the usual decimation and leaving the structure invariant. 
The key point is that here $\rho(\omega)$ and 
$P_{ii}(t)$ transform according to different laws under this new 
renormalization procedure.
This leads to independent asymptotic behavior for these quantities and requires
the definition of two distinct spectral dimensions 
we will call $\widetilde d_D$ and $\widetilde d_V$,
$D$ staying for {\it diffusive} and $V$ for {\it vibrational}.

\section{Harmonic oscillations and random walks on generic networks}
The harmonic oscillations of a generic network
of masses $m$ connected by springs of elastic constant $K$ are described
by the equations of motion for the displacements $x_i$ of each mass
from its equilibrium position:
\begin{equation}
m {d^2 \over dt^2} x_i = -K {\sum_{j \sim i}} (x_i-x_j)
\label{1}
\end{equation}
where the sum runs over the nearest neighbors of point $i$.\\
Equations (\ref{1}) can be Fourier transformed with respect 
to the time giving:
\begin{equation}
- {\omega^2 \over \omega_0^2} \tilde{x}_i = {\sum_{j \sim i}}
(\tilde{x}_j-\tilde{x}_i)
\label{2}
\end{equation}
where $\omega_0^2\equiv K/m$.\\
The equations describing random walks and harmonic oscillations
on a network are formally similar.
Let us consider a discrete time random walk (the so called
{\it blind ant} problem) on a network; the master equation for the probability 
of being at site $i$ after $t$ steps for a random walker starting from an 
origin site $O$ at time $0$, is:
$$
P_{Oi}(t+1) - P_{Oi}(t)= 1/z_{max} \sum_{j \sim i} (P_{Oj}(t) - P_{Oi}(t))
$$
\begin{equation}
P_{Oi}(0)=\delta_{iO}
\label{3}
\end{equation}
where $z_{max}$ is the maximum coordination number of the network (i.e. the 
maximum number of nearest neighbors of a point), that
is assumed to be finite.
Applying the  discrete Laplace transform with respect to the time 
defined by
\begin{equation}
\widetilde P_{ij}(\epsilon) = \sum_{t=0}^{\infty}(1+\epsilon)^{-t} P_{ij}(t)
\label{4}
\end{equation}
and taking $\epsilon\to 0$, the system (\ref{3}) can be written in a
form similar to (\ref{2}):
\begin{equation}
z_{max}\epsilon\widetilde P_{Oi}(\epsilon) =\sum_{j \sim i}
(\widetilde P_{Oj}(\epsilon)-\widetilde P_{Oi}(\epsilon))+\delta_{Oi}.
\label{5}
\end{equation}
After the substitutions $- z_{max}\epsilon \rightarrow {\omega^2/ \omega_0^2} 
=\gamma$ and $\widetilde P_{Oi}(\epsilon)\rightarrow \widetilde{x_i}$,
equations (\ref{2}) and (\ref{5}) look very similar. 
However they present a fundamental difference consisting in the term
$\delta_{Oi}$, arising from the initial condition in (\ref{3}).
In addition, the system  (\ref{2}) is homogeneous and defines an eigenvalue
problem which  has infinite solutions (all the normal modes of the graph). The
density of eigenvalues for $\omega\to 0$  depends on $\widetilde d_V$ and it is
not a direct solution of the system, but has to be separately calculated after
solving the whole system.
On the other hand, the system (\ref{5}) is inhomogeneous and corresponds
to a Cauchy problem, which has only one solution. The behavior of such a 
solution
for  $\epsilon\to 0$ depends on $\widetilde d_D$ and, in principle, has no
direct relation with the density of vibrational modes and with $\widetilde 
d_V$.
A relation between the random walk probabilities $P_{ii}(t)$ and $\widetilde
d_V$  does indeed exist but it involves the average of the $P_{ii}(t)$ over all
points of the graph \cite{prl}: 
\begin{equation}  
\bar{P}(t) = \lim_{N\rightarrow\infty} {1\over N} \sum_{i=1}^N P_{ii}(t)
\sim t^{-\widetilde d_V /2} ~~~~~\textstyle{ for }~~t\to\infty
\label{6}
\end{equation}
Relation (\ref{6}) does not imply that $\widetilde d_D = \widetilde d_V$
since the average $\bar{P}(t)$ has in general a different asymptotic behavior 
from each $P_{ii}(t)$.\\

\section{ Random walks and harmonic oscillations on exactly decimable fractals}

Exactly decimable fractals are a restricted class of self similar structures
(i.e. not all self similar structures are exactly decimable)
which are geometrically invariant under site decimation.
Examples of exactly decimable fractals are the Sierpinski Gasket 
\cite{ret}, the $T-$fractal \cite{ses}, the branched Koch curves \cite{koch}
and so on (Fig.1).
The solution of both the random walks and the harmonic oscillations problems
can be obtained by standard renormalization group calculations 
based on a real space decimation procedure \cite{giac}.\\
A geometrical structure is decimation invariant if it is possible to
eliminate a subset of points (and all the bonds connecting these points) 
obtaining a network with the same geometry of the starting one. 
From a mathematical point of view this 
corresponds to the possibility of eliminating by substitution a set of equations
from system (\ref{2}) or (\ref{5}) obtaining a system which is similar to the 
initial one after a suitable redefinition of the coupling parameter $\gamma$.
If we consider for example a $T-$fractal the decimation procedure consists in 
transforming each ``T" made of three bonds in a simple bond connecting two 
points. As it can be easily verified this operation does not change the geometry 
of the network but requires a redefinition of the coupling parameter $\gamma$.
In general, after a decimation step, $\gamma$ splits into
a finite number of different couplings $\gamma_\mu,~\mu=1,...,n$ on
geometrically inequivalent points. 
In the case of the $T-$fractal there are two kinds of inequivalent points:
points having one nearest neighbor and points with three nearest neighbors.
This suggests to distinguish between a coupling $\gamma_1$
and a coupling $\gamma_3$ to be used in equations (\ref{2}) or (\ref{5})
where point $i$ has respectively one or three nearest neighbors.
Before the decimation we have obviously $\gamma_1=\gamma_3=\gamma$ but this 
distinction is useful to put into evidence the forthcoming splitting.\\
The splitting of $\gamma$ in an at most finite number
of couplings is a necessary condition for exact decimability.
If this condition is fulfilled, the decimation transform can be iterated 
and linearized near the fixed point $\gamma_\mu = 0$. 
After the linearization the transformation laws for the $T-$fractal are
\[ \gamma_1\rightarrow \gamma_1' =3\gamma_1+1\gamma_3
\]
\begin{equation}
\gamma_3\rightarrow \gamma_3' =3\gamma_1+5\gamma_3
\end{equation}
The linearized decimation transform can be represented by a
matrix $D$ acting on a vector with components $\gamma_\mu,~\mu=1,...,n$
so that:
\begin{equation}
\gamma_\mu'=\sum_\nu D_{\mu \nu} \gamma_\nu
\end{equation}
In our particular case:
\begin{equation}
D=\pmatrix{ 3 & 1 \cr
 3 & 5 \cr}
\end{equation}
Decomposing $\gamma_\mu$ on the basis of eigenvectors of $D$,
as the number of decimation steps goes to $\infty$, $\gamma'_\mu$ tends to
the eigenvector corresponding to the largest 
eigenvalue of $D$.
Therefore the largest eigenvalue of $D$, $a^2$, determines the transformation
laws of coupling parameters.\\ 
Following these steps in the case of random walks one finds for the 
parameter $\epsilon$ the transformation:
\begin{equation}
\epsilon \rightarrow \epsilon'(\epsilon)\sim a^2\epsilon
\label{8}
\end{equation}
where $\epsilon$ is now the projection of the vector $\epsilon_\mu$ on the
largest eigenvalue direction.
The presence of the term $\delta_{iO}$ in (\ref{5}) requires a 
redefinition of the quantities $\widetilde P_{ij} (\epsilon)$ to assure that, 
even after the decimation, the initial condition will correspond to the 
probability of being in a fixed site equal to 1.
One introduces a new parameter $c$ and writes the transformation 
law for $\widetilde P_{ij} (\epsilon)$ as:
\begin{equation}
\widetilde P_{ij} (\epsilon) \rightarrow \widetilde P'_{ij} (\epsilon')
\sim {1\over c} \widetilde P_{ij} (\epsilon)
\label{9}
\end{equation}
The diffusive spectral dimension  $\widetilde d_D$ is obtained using the 
relation:
\begin{equation}
\widetilde P_{OO} (\epsilon) \sim \epsilon ^{ \widetilde d_D/2 -1}
\label{10}
\end{equation}
which holds only for $\widetilde d_D<2$. As we will discuss later, 
this is always the case for exactly decimable fractals.
Now one can rewrite (\ref{9}) as:
\begin{equation}
\widetilde P_{ij} (\epsilon) \sim c \widetilde P'_{ij} (\epsilon')
\end{equation}
and observe that since our graphs are infinite 
$\widetilde P_{OO} (\epsilon)$ and $ \widetilde P'_{OO} (\epsilon')$
refer to the same structure and 
they must have the same functional form. This gives: 
\begin{equation}
\epsilon^{\widetilde d_D/2 -1}=c(a^2\epsilon')^{\widetilde d_D/2 -1}
\label{11}
\end{equation}
so that:
\begin{equation}
\widetilde d_D= 2 {{ \log a^2/c}\over {\log a^2}}
\label{12}
\end{equation}
As for harmonic oscillations, the invariance of the 
network under the decimation procedure gives for $\omega ^2$ the same 
transformation law obtained for $\epsilon$. The analogous of (\ref{8}) 
is here:
\begin{equation}
\omega \rightarrow \omega '(\omega) \sim a \omega 
\label{13}
\end{equation}
The relation between the density of vibrational modes $\rho (\omega)$ 
and the new $\rho' (\omega')$ is given by:
\begin{equation}
\rho (\omega) d\omega = \rho' (\omega')d\omega'
\label{14}
\end{equation}
Since in the decimation procedure the initial set of equations 
(\ref{2}) is reduced of a factor $r$, condition (\ref{14}) 
leads to the relation:
\begin{equation}
\rho (\omega) \rightarrow \rho' (\omega') \sim {r\over a} \rho (\omega)
\label{15}
\end{equation}
From (\ref{15}) it follows:
\begin{equation}
\rho (\omega)= {a\over r} \rho' (\omega') 
\end{equation}
so that:
\begin{equation}
\omega^{\widetilde d_V -1} = {a\over r} (a \omega)^{\widetilde d_V -1}
\label{16}
\end{equation}
and:
\begin{equation}
\widetilde d_V ={ {\log r}\over {\log a}}
\label{17}
\end{equation}
Comparing (\ref{12}) and (\ref{17}) one realizes that 
$\widetilde d_V= \widetilde d_D$ if the decimation ratio $r$ is given by:
\begin{equation}
r=a^2/c
\label{18}
\end{equation}
This can be shown to be the case for exactly decimable fractals, using
results obtained \cite{hhw} for the Gaussian 
model on exactly decimable fractals. The Gaussian model 
on a graph is defined by the Hamiltonian:
\begin{equation}
H(\{m_i\}) = {J\over 4} \sum _{i\sim j} \left( \phi_i - \phi_j \right)^2+
\sum_i m_i ^2 \phi_i ^2
\label{19}
\end{equation}
The autocorrelation functions of the model are related to the 
generating functions $\widetilde P_{ii} (\epsilon)$ of random walks by:
\begin{equation}
\widetilde P_{ii} (\epsilon)=< \phi_i  \phi_i >_{\{ m_i ^2\}}\cdot 
{{z_i}\over {1-\epsilon}}
\label{20}
\end{equation}
where $z_i$ is the coordination number of site $i$ and the masses $ m_i ^2$
are given by $ m_i ^2= z_i \epsilon/(1-\epsilon)$.
Now it can be shown that a decimation procedure implies 
the following scaling relations 
for the coupling $J$ and the masses $m_i$:
\begin{equation}
J \rightarrow \alpha J\,\,\,\,\,\,\,\,\,\,\, 
m_i \rightarrow \beta m_i
\label{21}
\end{equation}
where the parameters $\alpha$ and $\beta$ are related to the expression of the
diffusive spectral dimension by:
\begin{equation}
\widetilde d_D = {{2 \log \beta} \over {\log \beta /\alpha }}
\label{22}
\end{equation}
Now the scaling relations (\ref{21}) can be rewritten as scaling relations for 
the fields $\phi _i$ and the masses $m_i$:
\begin{equation}
\phi _i \rightarrow \sqrt{\alpha} \phi _i\,\,\,\,\,\,\,\,\,\,\,
m_i \rightarrow {\beta \over \alpha}  m_i
\label{23}
\end{equation}
In terms of random walks the transformation on the masses in (\ref{23}) 
does not affect 
the asymptotic behavior of $\widetilde P_{ii} (\epsilon)$ \cite{prl}
while the first of (\ref{23}), thanks to relation (\ref{20}),
implies that $\widetilde P_{ii} (\epsilon)\rightarrow \alpha \widetilde P_{ii} 
(\epsilon)$ so that we can identify $\alpha$ and $1/c$. Since the parameter
$\beta$ is the decimation ratio $r$, from (\ref{22}),
(\ref{17}) and (\ref{12}) equality (\ref{18}) follows.\\
Notice also that since it is always $\beta >\alpha \neq 1$ \cite{hhw},
exactly decimable fractals have $\widetilde d_D= \widetilde d_V <2$.
From (\ref{18}), the dimensional degeneration $\widetilde d_D = 
\widetilde d_V$ can be obtained knowing only two out of the three parameters 
$a$, $r$ and $c$.
For example the Sierpinski Gasket has $r=3$ and $a=\sqrt 5$, the $T-$fractal
has $r=3$ and $a=\sqrt 6$.

\section{Random walks and harmonic oscillations on non exactly decimable
fractals: $NT_D$ and the cutting-decimation renormalization transform}

In the previous section we have analyzed spectral dimension 
degeneration on exactly decimable fractals.
Now we will consider the case of a class of fractal graphs ($NT_D$) which are 
invariant under a more complex 
transformation $T=D\cdot C$ consisting of the product of 
a cutting transform $C$ and a decimation $D$.
We call $T$ a cutting-decimation transform. 
It will be shown that $\widetilde P_{ii} (\epsilon)$ and $\rho (\omega)$
behave differently under $T$ and that $NT_D$ are an explicit 
example of fractals with different diffusive and vibrational 
spectral dimensions.\\
The fractal trees known as $NT_D$ \cite{doyle} can be recursively defined as
follows:  an origin point $O$ (Fig.2)
is connected to a point $1$ by a link, of unitary 
length; from $1$, the tree splits in $k$ branches of length $2$ (i.e. 
consisting of two consecutive links); the ends of these
branches split in $k$ branches of length $4$ and so on; each
endpoint of a branch of length $2^n$ splits in $k$ branches of length $2^{n+1}$.
$NT_D$ can be naturally embedded in a suitable Euclidean space in such a way 
that their fractal dimension $d_F$ coincides with their connectivity dimension
$d_C = 1 + \ln k/ \ln 2$ \cite {der}.\\
As one can easily verify, $NT_D$ are not exactly decimable since after a
simple decimation starting from the origin $O$, one obtains $k$ copies of the
original structure joined together in a point instead of the same $NT_D$.
The $NT_D$ invariance under a $T$ transform can be described in the 
following way (Fig.3).
Suppose to cut the log of the tree in point $1$ and to separate the $k$ 
branches (cutting transform). Now, each branch can be obtained
from the initial $NT_D$ by a dilatation with a factor 2.
Eliminating all branches but one and decimating it 
(decimation transform), one obtains the original $NT_D$. 
The $T$ transform can be applied to solve the random 
walks problem. 
The cutting transform gives a  relation between random
walks on the whole tree and random walks on one of its branches; more precisely 
one relates $\widetilde P_{OO} ^{tree} (\epsilon)$, the generating function of 
the probability of returning to point $O$ after a random walk on the $NT_D$ 
tree, and $\widetilde P_{11} ^{branch} (\epsilon)$, the generating function of
the probability of returning to the starting point $1$ after a random walk on
one  of the branches. This relation is given by \cite{der}:
\begin{equation}
\widetilde P_{OO} ^{tree}(\epsilon) ={ {\widetilde P_{11} ^{branch} (\epsilon)
+k} \over {2\epsilon \widetilde P_{11} ^{branch} (\epsilon) +k}}
\label{24}
\end{equation}
The decimation transform on the surviving branch is obtained 
as for exactly decimable fractals by the transformation laws:
$$
\epsilon \rightarrow a^2 \epsilon= 4 \epsilon
$$
\begin{equation} \widetilde P_{11} ^{branch} (\epsilon) \rightarrow \widetilde
{P'}_{11} ^{branch}
(\epsilon')={1\over c} \widetilde P_{11} ^{branch} (\epsilon)= 
{1\over 2} \widetilde P_{11} ^{branch} (\epsilon)
\label{25}
\end{equation}
Now since the decimation procedure transforms a branch into the initial tree
we have 
$\widetilde {P'}_{11} ^{branch}(\epsilon')= \widetilde P_{OO}^{tree}
(\epsilon')$ and relation (\ref{24}) becomes:
\begin{equation}
\widetilde P_{OO}^{tree}(\epsilon)={  { 2\widetilde P_{OO}^{tree}(4\epsilon)
+k} \over { 4\epsilon \widetilde P_{OO}^{tree}(4\epsilon) +k}}
\label{26}
\end{equation}
Choosing a suitable expression for $P_{OO}^{tree}(\epsilon)$ \cite{der}
we obtain from (\ref{26}):
\begin{equation}
\widetilde d_D = 1+ {{\log k} \over {\log 2}}
\label{27}
\end{equation}
As for harmonic oscillations, the cutting transform gives a 
relation between $\rho ^{tree} (\omega)$ and $\rho ^{branch} (\omega)$;
this can be obtained using the following properties \cite{momo}:
1) cutting or adding a finite number of points or bonds from an infinite 
network does not affect its vibrational spectrum; 2)
the spectral density $\rho(\omega)$ (normalized to 1) 
of $k$ copies of a given structure all attached in a point coincides with the
corresponding one for a single structure.
From these properties it immediately follows that the 
cutting transform does not affect the density of modes $\rho (\omega)$ 
so that:
\begin{equation}
\rho ^{tree} (\omega)= \rho ^{branch} (\omega)
\label{28}
\end{equation}
Applying the decimation transform we get $a=2$ from (\ref{25}) 
while the decimation ratio is $r=2$ so that (\ref{15})
becomes:
\begin{equation}
\rho ^{branch} (\omega) \rightarrow {\rho '}^{branch} (\omega ')\sim 
\rho ^{branch} (\omega) 
\label{29}
\end{equation}
Since ${\rho '}^{branch} (\omega ')= \rho ^{tree} (\omega ')$ relations
(\ref{28}) and (\ref{29}) give us $\rho ^{tree} (\omega) =\rho ^{tree} 
(\omega')$ so that $\widetilde d_V =1$.\\
Since $\widetilde d_D = 1 + \log k /\log 2$ and $\widetilde d_V =1$,
$NT_D$ trees represent an explicit case of dynamical dimension splitting. 
All our results can be generalized 
\cite{albert} to $NT_D$ lattices with a branches growth factor $r$
(or decimation ratio) different from 2. In all these cases we will 
find again dynamical dimension 
splitting since $\widetilde d_V=1$ and:
\begin{equation}
\widetilde d_D =\widetilde d_V + {  {\log k} \over {\log r}}
\end{equation}
\section{Conclusions}

In this paper we have shown how to built an exact renormalization group 
technique in real space using a more general reduction procedure
than the usual decimation transform.
This new technique allows to point out the phenomenon of dynamical 
dimension splitting which, as we have explicitly shown, is always absent in the 
case of exactly decimable fractals, where
$\widetilde d_D= \widetilde d_V <2$. Therefore these structures have 
very peculiar statistical mechanical properties and their study
does not allow to distinguish between average and local quantities.

Many examples of spectral dimension splitting can be found even on non fractal
graphs, such as comb lattices \cite{pettini} and other
branched structures. In general the situation with $\widetilde d_D
\neq  \widetilde d_V$ is the most common so that it becomes 
particularly important to study the consequences of dynamical dimension 
splitting in all the phenomena which are influenced by the value of the 
spectral dimension such as diffusion, vibrational dynamics and 
phase transitions.

\vspace{1cm}
\centerline{Figure captions}
\vspace{1cm}
\noindent{\bf Fig.1}  Exactly decimable fractals: 
a) Sierpinski Gasket, b) $T-$fractal
c) Branched Koch curves\\
  \\
{\bf Fig.2} $NT_D$ with $k=3$\\
  \\
{\bf Fig.3} Cutting-Decimation procedure:\\
a) Cutting of the log of the $NT_D$\\
b) Separation of the $k$ branches\\
c) Decimation of the points labelled by $X$\\
d) Recovering of the original $NT_D$\\


\begin{thebibliography}{99}

\bibitem{aeo}
S.Alexander and R.Orbach, {\it Journal de Physique -- Lettres}, {\bf 43}, L625,
(1982)
\bibitem{hhw}
K.Hattori, T.Hattori and H.Watanabe, {\it Progr. of Theor.
Phys. Suppl.}, {\bf 92}, 108, (1987)
\bibitem{prl}
R. Burioni and  D. Cassi {\it Phys. Rev. Lett.}, {\bf 76}, 1091, (1996)
\bibitem{r+d+s}
R.Burioni, D.Cassi and S. Regina, {\it Modern
Physics Letters B}, {\bf 10}, 1059, (1996)
\bibitem{ret}
R.Rammal and G.Toulouse, {\it Journal de Physique -- Lettres}, {\bf 44}, L13, 
(1983)
\bibitem{ses}
A.Maritan, G.Sartoni and A.L.Stella, {\it Physical Review Letters}, {\bf
71}, 1027, (1993)
\bibitem{koch}
Y.Gefen, A.Aharony and B.B.Mandelbrot, {\it Physical Review Letters}, {\bf
45}, 855, (1980)
\bibitem{giac}
A.Giacometti, A.Maritan and A.L.Stella, {\it International
Journal of Modern Physics B}, {\bf 5}, 709, (1991)
\bibitem{doyle}
P.G. Doyle and J.L. Snell, {\it Random Walks and Electric Networks},           
(The Mathematical Association of America, 1984)
\bibitem{der}
R.Burioni and D.Cassi, {\it Physical Review E}, {\bf 51}, 2865, (1995)
\bibitem{momo}
B.Mohar, {\it Lin. Alg. and Appl.}, {\bf 48}, 245, (1982)
\bibitem{albert}
R.Burioni, D.Cassi, A.Pirati and S.Regina, preprint UPRF-97-10
\bibitem{pettini}
G.H.Weiss and S.Havlin, {\it Physica}, {\bf 134A}, 474, (1986)
\end{thebibliography}
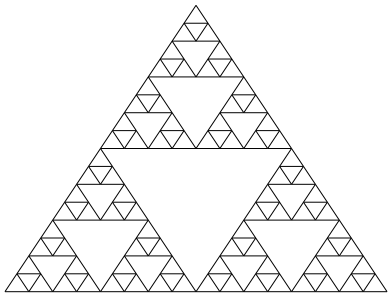
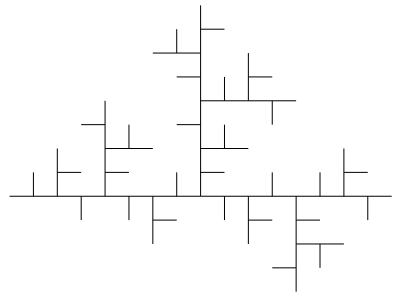
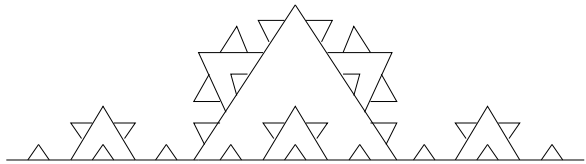
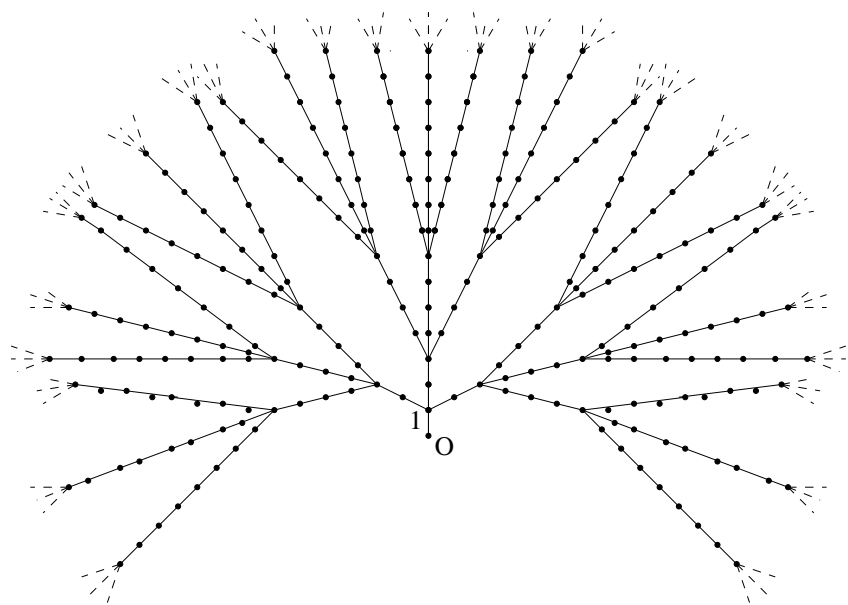
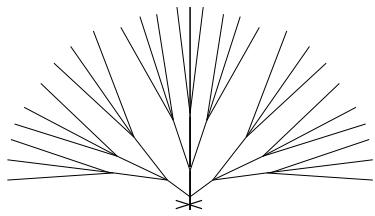
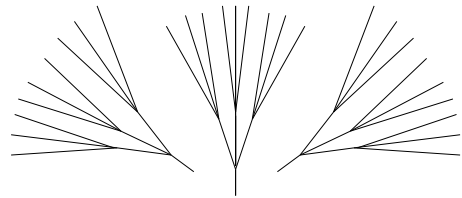
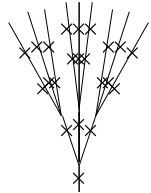
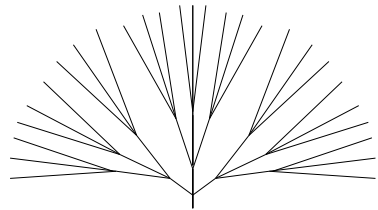
\end{document}